\documentstyle[eqsecnum,twocolumn,aps,prl,epsfig]{revtex}
\date{\today}
\draft
\tighten
\begin{document}
\def\sqr#1#2{{\vcenter{\hrule height.3pt
      \hbox{\vrule width.3pt height#2pt  \kern#1pt
         \vrule width.3pt}  \hrule height.3pt}}}
\def\square{\mathchoice{\sqr67\,}{\sqr67\,}\sqr{3}{3.5}\sqr{3}{3.5}}
\def\today{\ifcase\month\or
  January\or February\or March\or April\or May\or June\or July\or
  August\or September\or October\or November\or December\fi
  \space\number\day, \number\year}

\def\Bbb{\bf}


\title{Abelian Higgs hair for extreme black holes and selection rules 
for snapping strings}

\author{A. Chamblin{$^{1,4}$}, J.M.A. Ashbourn-Chamblin{$^{2}$},
R. Emparan{$^3$}, A. Sornborger{$^4$}}

\address {\qquad \\ {$^1$}Institute for Theoretical Physics,
University of California,
Santa Barbara, CA 93106-4030, U.S.A.
\qquad \\ {$^2$}Wolfson College,
University of Oxford,
Oxford OX2 6UD, England
\qquad\\{$^3$}Dept. of Physics,
University of California,
Santa Barbara, CA 93106, U.S.A.
\qquad\\{$^4$}DAMTP,
Silver Street,
Cambridge, CB3 9EW, England
}
\maketitle

\begin{abstract}
It has been argued that a black hole horizon can support 
the long range fields of a Nielsen-Olesen string, and that one 
can think of such a vortex as black hole ``hair''. 
We show that 
the fields inside the vortex
are completely expelled from a charged black hole in the extreme limit
(but not in the near extreme limit).  
This would seem to imply that a vortex 
cannot be attached to an extreme black hole. Furthermore, we provide 
evidence that it is energetically unfavourable for a thin vortex 
to interact with a large extreme black hole. This 
dispels the notion that a black hole can support `long' Abelian Higgs 
hair in the extreme limit. We
discuss the implications for strings that end at black holes, as in
processes where a string snaps by nucleating black holes.
\end{abstract}



Black hole `hair' is defined to be any field(s) associated with a stationary
black hole configuration which can be detected by asymptotic observers
but which cannot be identified with electromagnetic
or gravitational degrees of freedom.  
A number of results have been proven (\cite{israel})
which imply that black holes `have no hair'.  
These results led people to believe that
a black hole horizon can only support charges associated with long range gauge 
fields.  However, this prejudice was to some extent discredited
when Bartnik and McKinnon \cite{bart} discovered a solution of the Einstein-
Yang-Mills equations which supported Yang-Mills fields which can be detected 
by asymptotic observers; one therefore says that these black holes are 
{\it coloured}.
However, these exotic solutions do not impugn the original no hair results
since all such solutions are known to be unstable (see e.g. \cite{biz}).
Since the original no hair theorems assumed a stationary
picture they simply do not apply to coloured holes. On the other hand,
coloured holes do still exist and so they are said to `evade' the usual
no hair results.
These results teach us that we have to tread carefully when
we start talking about black hole hair. 

With this in mind, we analyze the extent to which hair
is present in situations involving topological defects, such as cosmic strings 
\cite{review}. 
In \cite{ana} evidence was presented that a 
Nielsen-Olesen (Abelian) vortex can `thread' a Schwarzschild
black hole. Inclusion of the gravitational back reaction of a single 
thin vortex led to a metric 
which is just a conical defect centered on a black hole (\cite{aryal}). 
Thus, it was argued that 
this solution truly is the `thin vortex' limit of a `physical' 
vortex-black hole configuration.  Given these results, one
can conclude \cite{ana} that the Abelian Higgs vortex is
{\it not} just dressing for the Schwarzschild black hole, but rather that 
the vortex is truly hair, i.e., a property of the black hole which can
be detected by asymptotic observers.

In this paper, we extend the analysis of \cite{ana} and allow the black hole
to be charged.  That is, we consider the problem of an Abelian Higgs
vortex in the Reissner-Nordstrom background.  In order to `turn up' the 
electric charge of the hole, we have to allow for the presence of two
$U(1)$ gauge fields (one $U(1)$ is where the charge of the hole lives and 
the other $U(1)$ is the symmetry spontaneously broken in the ground state). 
We find two striking phenomena:

(i) In the extreme limit (but not near extremality)
{\it all} of the fields associated with the vortex 
(both the magnetic and scalar degrees of freedom) are expelled from 
the horizon of the black hole. The magnetic and scalar fields always 
`wrap around' the horizon in the extremal limit.  

(ii) By considering the total energy of the vortex with a black hole inside 
it, we find an 
instability as we allow the extreme black hole to become very large
compared to the size of the vortex. Specifically, the energy of a 
vortex which does {\it not} contain the hole inside it is {\it much} less 
than the energy of a vortex which does contain the hole.

In a sense, the behaviour (i) was expected, given that extreme black holes 
generically
display such a `Meissner effect', and so can be thought of as `superconductors'
(a deeper analysis of the superconducting properties of extremal black holes
and p-branes will be given in \cite{cham}).
But from (ii) it follows that a very thin vortex will want to `slide' off of 
the hole. Thus, the vortex cannot in any way be thought of as a 
`property of the black hole which can be measured at infinity'; 
in other words, an Abelian Higgs vortex is not hair for an extreme black hole. 

Our treatment of the black hole/string vortex system involves a clear
separation between the degrees of freedom of each of these objects. 
The action takes the form $S = S_1  + S_2$,
where $S_1$ is an Einstein-Hilbert-Maxwell action for the `background' 
fields $(g_{\mu\nu}, {\cal F}_{\mu\nu})$,
and $S_2$ describes an Abelian Higgs system minimally coupled
to gravity, 
\begin{equation}\label{abhiggs}
S_2 = \int d^4 x \sqrt{-g} \left( D_\mu\Phi^\dagger D^\mu\Phi -
{1\over 4 e^2} F^2 - {\lambda\over 4}(\Phi^\dagger\Phi - \eta^2)^2\right).
\end{equation}
The degrees of freedom in $S_2$ are treated as `test fields'.
They are the complex Higgs 
field, $\Phi$, and a $U(1)$ gauge field with strength, $F_{\mu\nu}$, 
and potential $A_\mu$. The Higgs scalar and the gauge field 
are coupled through the gauge covariant derivative
$D_\mu=\nabla_\mu + iA_\mu$, where $\nabla_\mu$ is
the spacetime covariant 
derivative. We choose metric signature $(+---)$.
It is also convenient to define the Bogomolnyi parameter
$\beta =\lambda/2e^2= m^2_{\rm Higgs}/m^2_{\rm vector}$. 
Notice that we have two different gauge fields: $F$, which
couples to the Higgs field and is therefore subject 
to spontaneous
symmetry breaking,
and ${\cal F}$, which remains massless.

A vortex is present when 
the phase of $\Phi (x)$ is a non-single valued quantity. 
To better describe
this, define the real fields $X$, $P_\mu$, $\chi$, by
$\Phi = \eta X e^{i\chi}$ and $A_\mu = P_\mu -\nabla_\mu\chi$.
The vortex is then characterized by 
$\oint d\chi = 2\pi N$, the integer $N$ being called 
the winding number. 

We will analyze the equations of the vortex in the background 
of the Reissner-Nordstrom black hole
\begin{eqnarray}
ds^2 &=& V dt^2 - {d\rho^2\over V} - \rho^2(d\theta^2 + 
\sin^2\theta d\varphi^2),\\
V &=& 1 - {2 G m\over \rho} + {q^2\over \rho^2}.\nonumber
\end{eqnarray}
We will work in rescaled coordinates and
parameters $(r,\;E,\;Q)=\eta\sqrt{\lambda}(\rho,\; G m,\;q)$. 
In these non-dimensional variables the Higgs mass is unity.
The Reissner-Nordstrom black hole has inner and 
outer horizons where $V(r)=0$. We will only be interested in the outer 
horizon, which is at radius
$r_+ = E +\sqrt{E^2-Q^2}$. If $r_+ = E = |Q|$,  then $V(r)$ has
a double zero at $r_+$, 
and the black hole is said to be extremal.

Return now to the equations of the vortex. One can consistently take
$X= X(r,\theta)$, $P_\varphi = N P(r,\theta)$,
which simplifies the equations of motion 
to the form
\vspace*{-2mm}
\begin{eqnarray}\label{rnoeq1}
&&-{1\over r^2}\partial_r( r^2 V\partial_r X) - {1\over r^2\sin\theta}
\partial_\theta(\sin\theta\partial_\theta X)\nonumber \\ 
&&+ {1\over 2} X(X^2 -1) 
+ {N^2 X P^2\over r^2\sin^2\theta} =0,
\end{eqnarray}
\begin{equation}\label{rnoeq2}
\partial_r(  V\partial_r P) + 
{\sin\theta\over r^2}\partial_\theta\left(
{\partial_\theta P\over \sin\theta}\right) - {X^2 P\over \beta} =0.
\end{equation}
When $P=1$ (a constant)
throughout the space we recover
a global string in the presence 
of the charged hole. 
The equations (\ref{rnoeq1}), (\ref{rnoeq2}), are, in general,
rather intractable in exact form and we need to resort to approximation 
methods. An analytical solution
of these equations for the case
where the black hole is small relative to the vortex size 
is constructed in \cite{caes}. Here we resort to numerical 
integration of
the equations (\ref{rnoeq1}) and (\ref{rnoeq2}) outside and 
on the black hole horizon.

The abelian Higgs equations in the presence of a background
Reissner-Nordstrom metric are elliptic. On the horizon they become
parabolic. In order to solve the equations numerically, we use a
technique first used by Ach\'ucarro, Gregory and Kuijken \cite{ana}. 
More details can be found in that reference and in \cite{caes}.
We have pushed this calculation to the limits, making the vortex as small
as we could given the computational constraints.  What we have found is that
the vortex is {\it always} expelled, no matter how small the magnetic and Higgs
flux tubes are taken to be.
Here we present a selection of dramatic pictures of the numerical 
evidence which we have amassed. The general pattern displayed
here holds no matter how small you make the flux tubes.

We begin with the expulsion of the $P$ field by the extreme hole. In the
diagram below, we have set $E = Q = 10$, with winding number $N = 1$ 
(the smallest winding possible). The Bogomolnyi parameter 
$\beta$ is set equal to unity, so that the magnetic and Higgs flux tubes 
are the same size:\\
\vspace*{0.1cm}

\epsfxsize=3cm
\epsfysize=6cm
\hspace*{1cm} \epsfbox{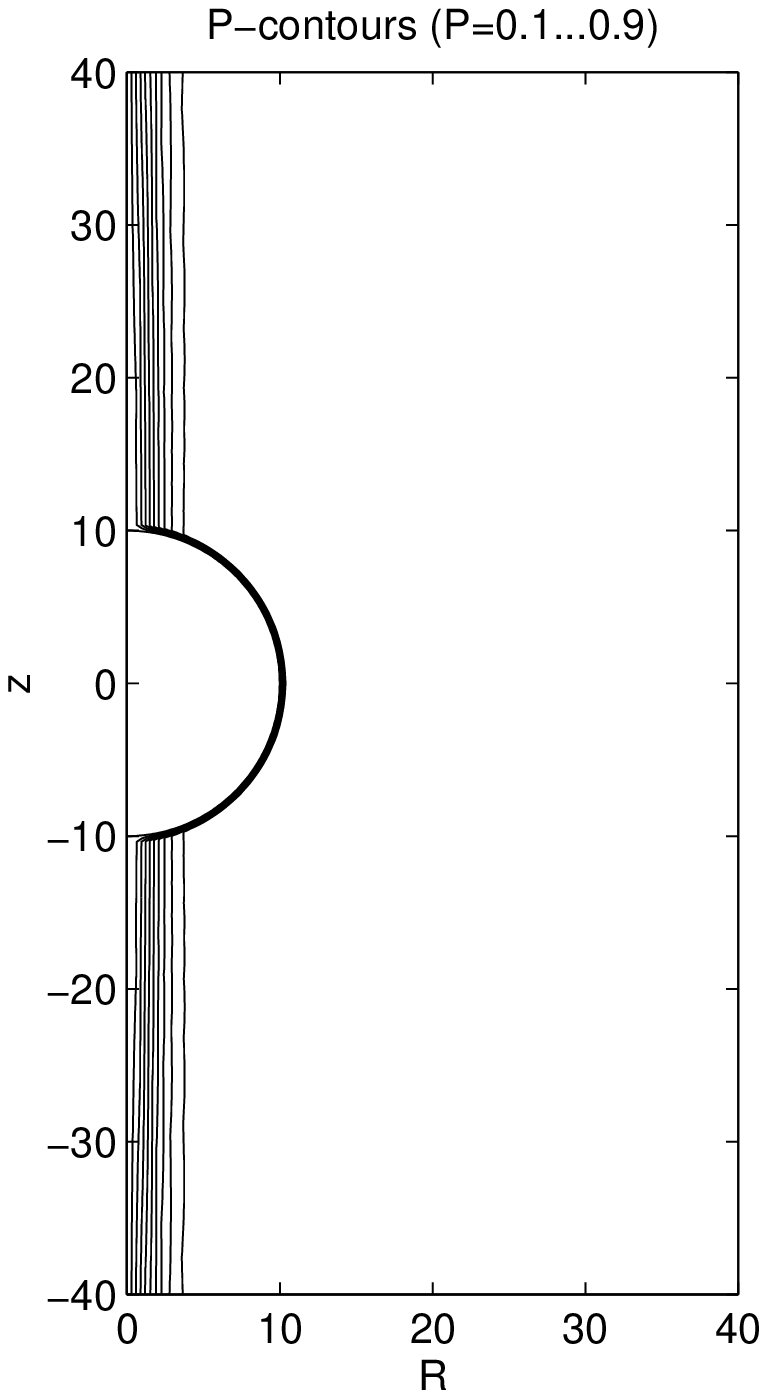}

{\noindent \small {\bf Figure 1:} Expulsion of the P field from the extreme 
horizon, for the values $E = Q = 10$, $N = 1$ and ${\beta} = 1$.}\\
\vspace*{0.1cm}

Clearly, the $P$ field `wraps' the black hole horizon; furthermore,
given the relation between $P$ and $F_{{\theta}{\varphi}}$ it is clear that 
no magnetic flux is crossing the horizon.  The extreme hole behaves
just like a perfect diamagnet.  But can we `puncture' the
horizon with flux by making the magnetic flux tube even smaller?  
The simplest way to make the vector flux tube thinner is by decreasing the
value of $\beta$. 
Since $\beta$ is the ratio of the sizes of the vector and
Higgs flux tubes, making $\beta$ very small corresponds to making 
the magnetic flux tube very skinny. 
However, we still find  the $P$ contours all wrap 
around the black hole horizon, indicating
that there is never any penetration.  

We now turn to the behaviour of the Higgs field $X$. 
We have found that the $X$ field is always
expelled from the extreme hole, no matter how small the scalar flux tube is
made.  Actually, in the figures below what we do is fix the
size of the scalar flux tube (by fixing $N = 1$ and ${\beta} = 0.5$) and we
allow the mass of the extreme hole to increase.  The plots run from left to right
with increasing mass.  The graphs are plotted for the values $E = Q = 1$,
$5$, $10$, and $20$.
\vspace*{0.2cm}

\centerline{\epsfxsize=2cm \epsfysize=4cm \epsfbox{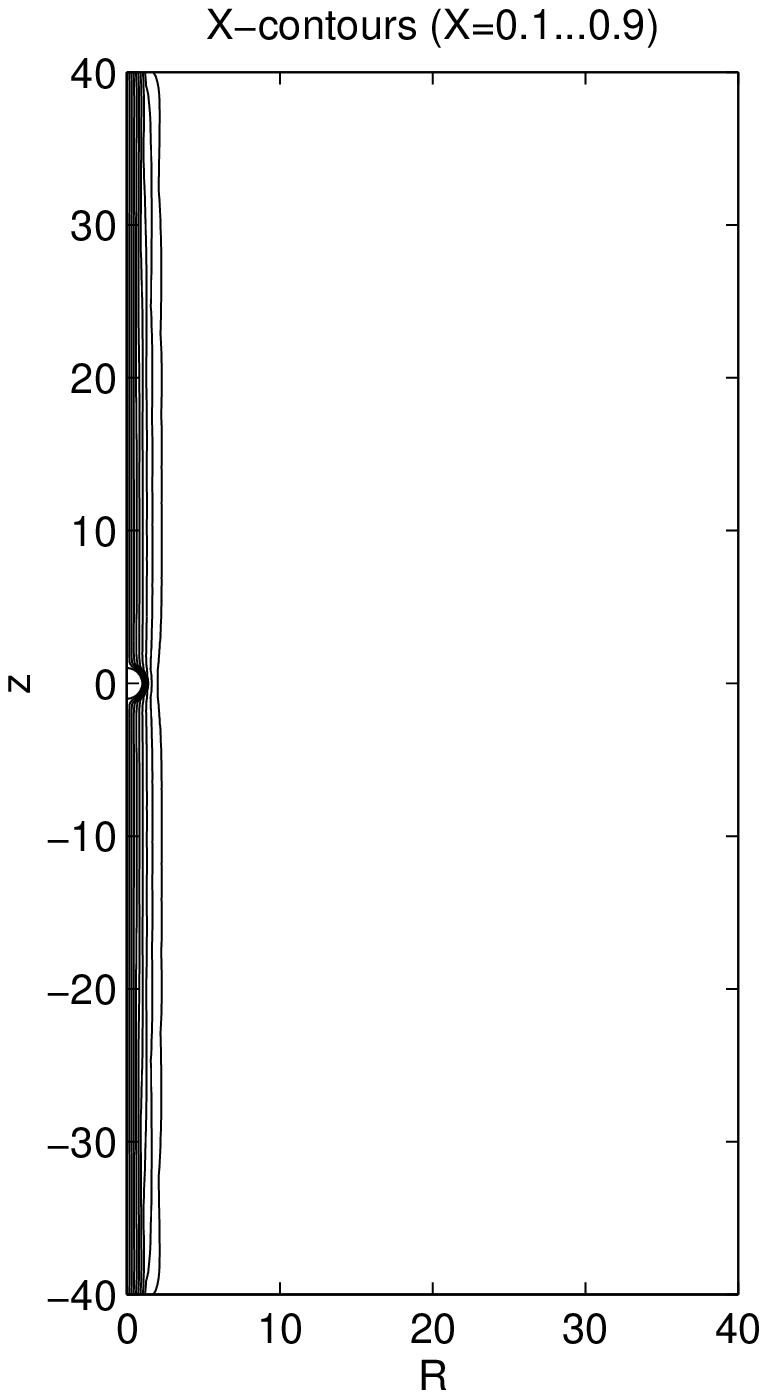}  
\epsfysize=4cm \epsfxsize=2cm 
\epsfbox{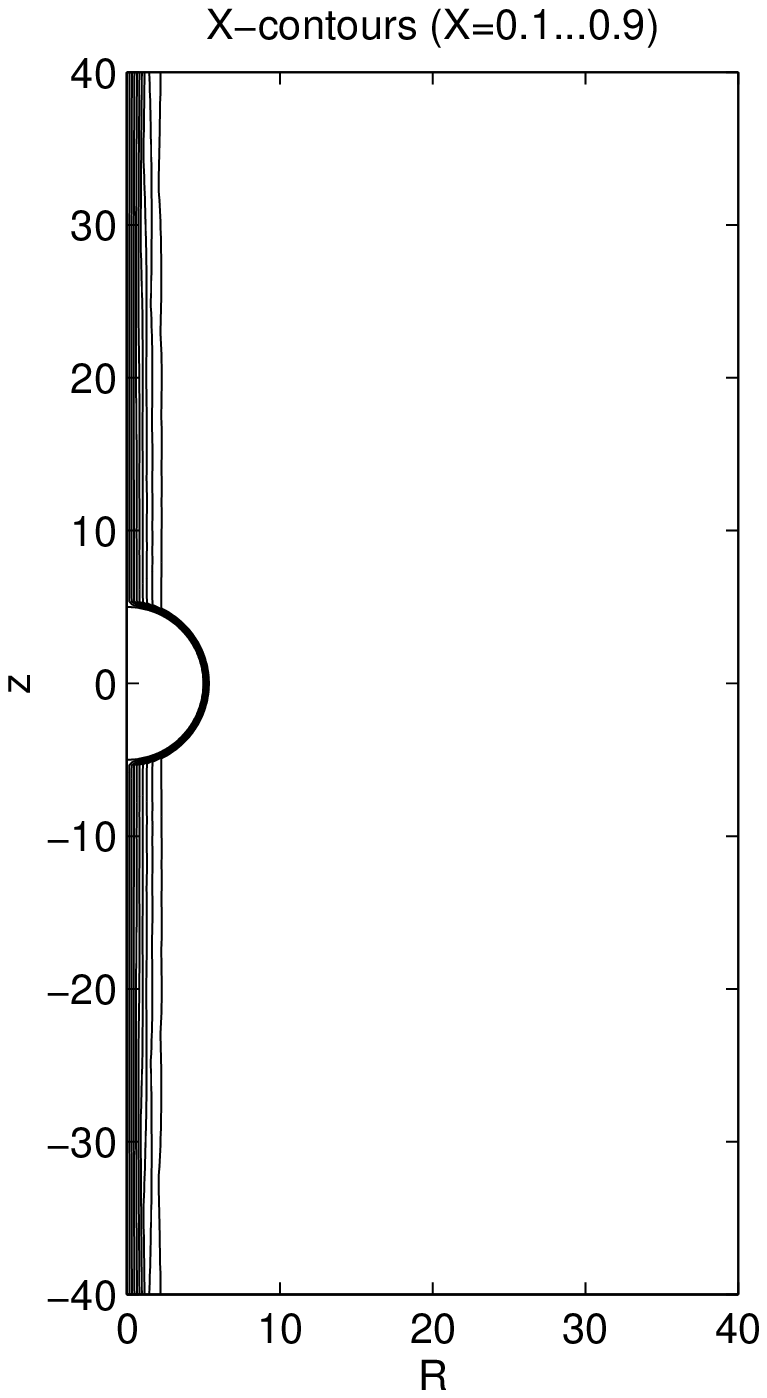} \epsfysize=4cm \epsfxsize=2cm \epsfbox{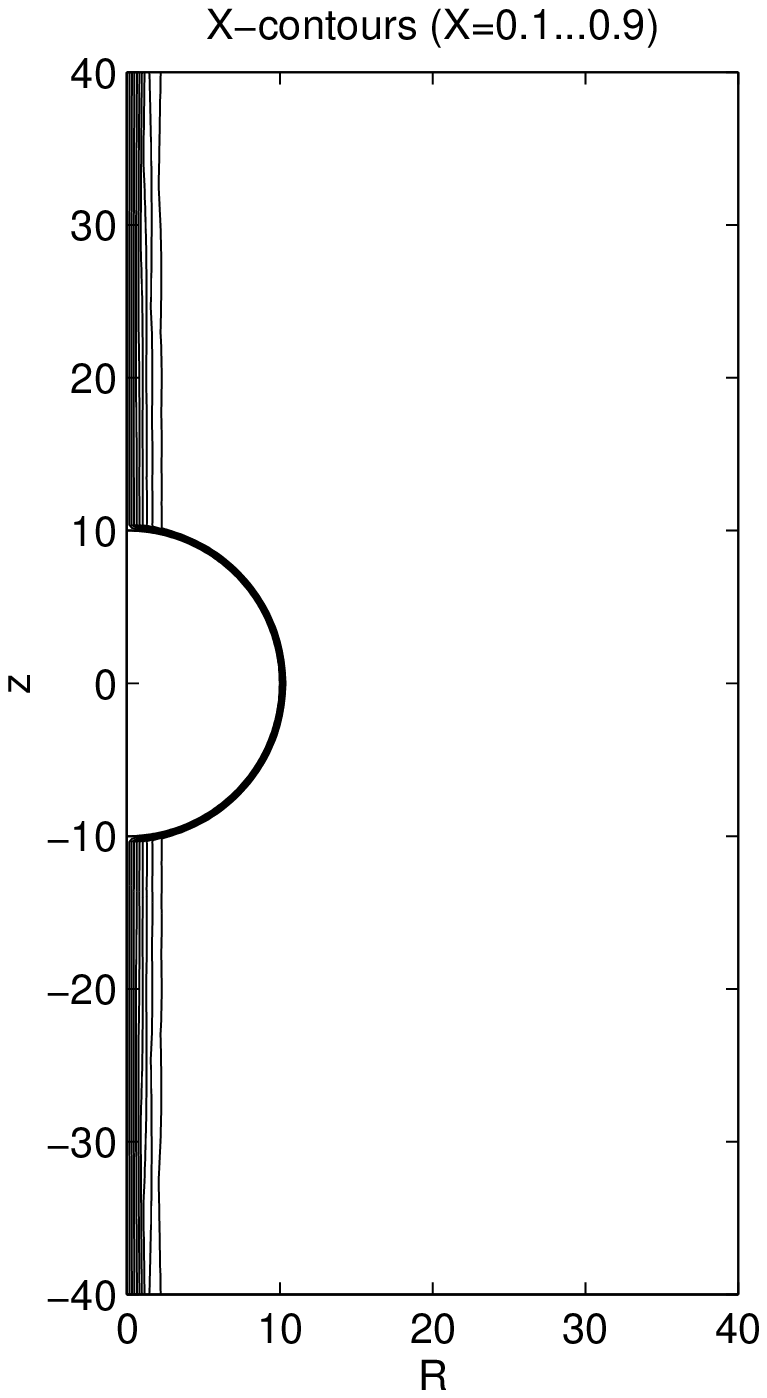}
\epsfysize=4cm \epsfxsize=2cm \epsfbox{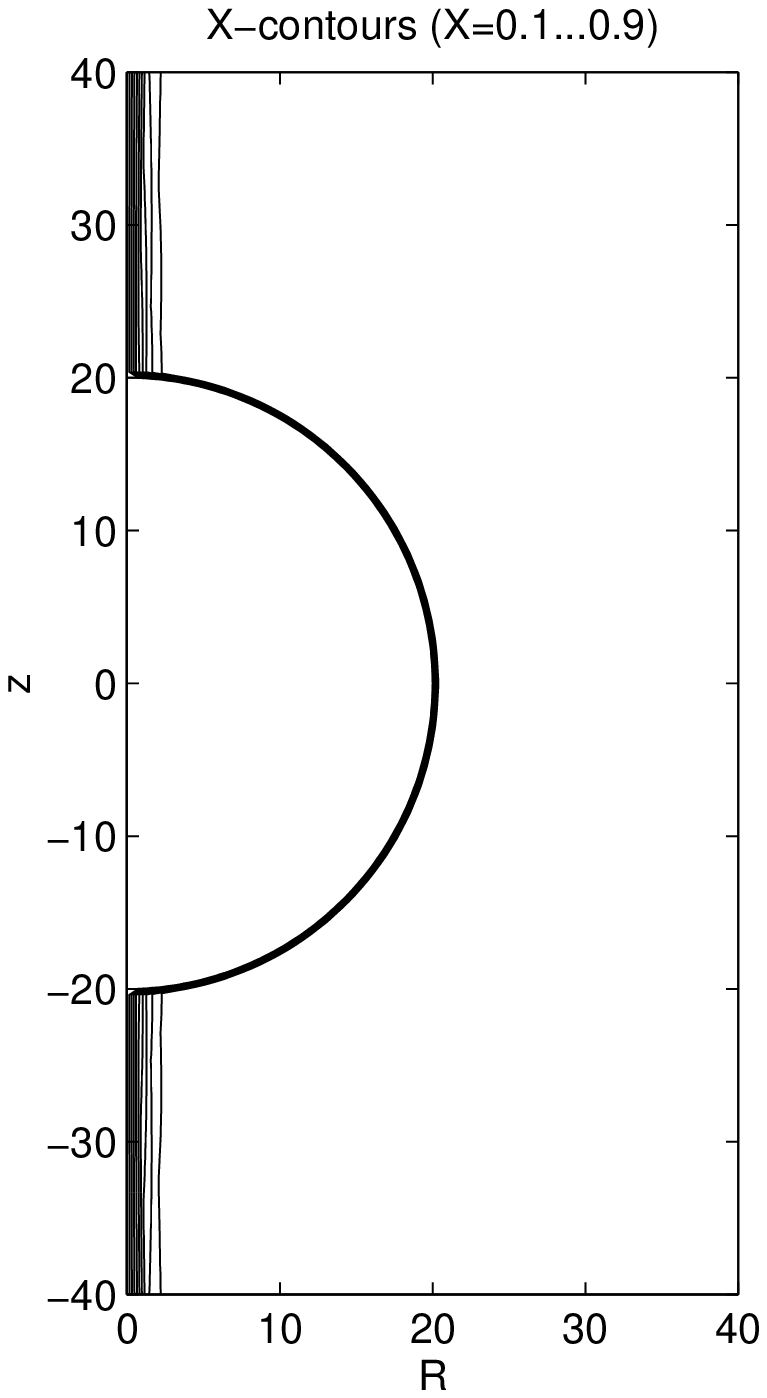}}
{\noindent \small {\bf Figure 2:} Expulsion of the Higgs field from the 
extreme horizon, for the values $E = Q = 1, 5, 10, 20$; $N = 1$ and 
${\beta} = 0.5$.}\\
\vspace*{0.1cm}

The $X$ contours all wrap around the black hole horizon,
no matter how large the hole is made. The effect is still true for global
strings, where the gauge
dynamics is absent.

Now consider the stability of the configurations. Is the black hole
stable inside the vortex, or will it try to find its way outside the core?
The above sequence provides an intuitive answer to this question.
When the black hole is much smaller than the vortex, the black hole is just a
`hole' where no vortex energy can be stored.  Thus, the presence of the hole
tends to subtract the total energy of the vortex.  On the other hand, when
the hole becomes much larger than the vortex,
flux stretches to wrap the hole and so we would expect the total energy
of the vortex to become very large.  

We have computed the
total energy ${\cal E}_{bh}$ stored in the vortex when it 
contains a black hole inside it,
for different relative sizes of the core and the horizon 
(note: in all numerical calculations we introduce an obvious
cutoff, i.e., we do not integrate over all of spacetime to obtain the energy, 
rather we integrate out to the boundaries of some large `box'). This is to 
be compared with the energy of 
the vortex in the absence of the black hole, ${\cal E}_{0}$. It is 
always the case that there exists a 
maximum mass $E_{max}$ such that for all black holes of mass $E < E_{max}$,
${\cal E}_{bh}(E) < {\cal E}_{0}$; as long as the hole is not 
too massive, it prefers to sit inside the vortex. 

The statements made above are based on the results of our numerical computations
of the total energy ${\cal E}_{bh}$. In the figure below we have plotted
the results of one such computation. Here, we have set ${\beta} = 0.5$ and
$N = 10$.  The flat, horizontal line (at 6640) represents ${\cal E}_{0}$
in our units.  
Clearly, for these values $E_{max}$ is about 15.
Furthermore, for black holes of mass greater than $E_{max}$ the energy of the
vortex is diverging.  The erratic behaviour of the vortex energy for very
small values of the black hole mass is an artifact of the numerical techniques
employed in the calculation and should be ignored:\\
\vspace*{0.1cm}

\epsfxsize=7cm
\epsfysize=7cm
\hspace*{0.1cm} \epsfbox{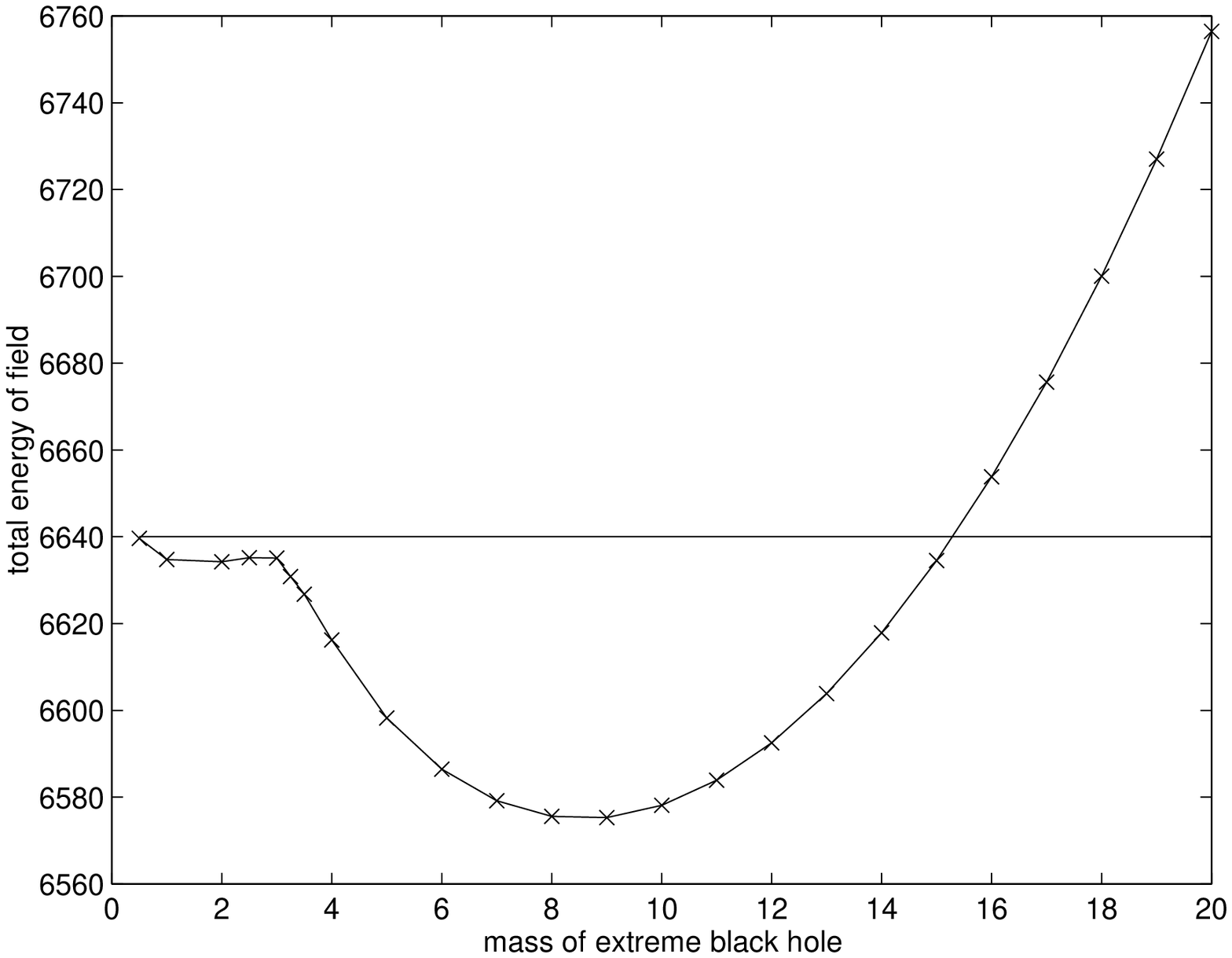}

{\noindent \small {\bf Figure 3:} Plot of total vortex field energy as a 
function of black hole mass.}\\
\vspace*{0.1cm}

It is clear that a black hole with mass $E > 15$ is going to
find it energetically favourable to slip out of the vortex.  Thus, it is 
not appropriate to think of such a vortex as a `property of the black hole';
the identification of the vortex as long hair does not go through
in this situation.  When the mass of the hole is small you could
still try to identify the vortex with hair since at least in that
case the configuration is energetically stable.  On the other hand, the fact
remains that the vortex is completely expelled from the hole, even in the 
(putatively) stable situation. Thus one would say that the vortex is not 
{\it dressing} the black hole. It is not clear to us whether or not 
one should think of such a `thick' vortex as genuine hair for a small 
extreme black hole. This is somewhat a reversal of previously studied 
situations (e.g., the coloured black holes), where the black hole may 
be dressed but the configuration is unstable.

A natural question is whether or not similar results continue to hold
when the hole is {\it slightly} non-extreme. Our numerical calculations
show that, even when the vortex is very
thin relative to the radius of the hole, and the charge is very close to 
extremality, the flux is expelled only in the exact extreme limit (see
\cite{caes} for more details).

We have provided strong evidence that the fields of a vortex are
always expelled from an extreme horizon. Furthermore, a 
thin enough vortex tends to slip off the black hole. Thus, it appears 
that an extreme black hole cannot support `long' Abelian Higgs hair.
Of course, we have not accounted for the
back reaction of the vortex on the geometry.
But there is evidence that the expulsion may hold 
exactly: there do exist {\it exact} solutions 
for black holes in $U(1)^2$ theories where a black hole that is
charged to extremality with respect to one of the gauge fields, completely
expels the field of a (Melvin) flux tube of the other gauge field 
\cite{cham}. This strongly suggests that, after 
accounting for backreaction, the flux should be expelled from an 
extreme black hole that sits inside it, at least when the 
vortex is thick. 
Given the evidence provided above, the 
effect could well persist in the thin vortex limit.
In any case, back reaction would have to be
small if the energy scale of symmetry breaking is small compared 
to the black hole mass.

If, as we have argued, vortices fail to penetrate 
extreme horizons, there are several interesting implications.
Consider what happens when a string tries to {\it end} at a black hole.
It has been argued in \cite{ana} that there is no global topological
obstruction for a topologically stable string to end at a black hole. 
This argument is still valid for extreme Reissner-Nordstrom black holes.
But what we have found seems to strongly suggest 
that, even if the penetration on only one side of the hole 
is topologically feasible, there does not
exist a solution of the equations of motion that actually penetrates. 
We have not analyzed the situation where the string is only on one 
side of the black hole, but
our results very strongly hint that there is no way a vortex can penetrate 
an extreme horizon: the penetration is 
a {\it local} issue, unrelated to global topological 
considerations. 
If the string cannot pierce the extreme horizon, then there is no way
to construct the Wu-Yang type of patch for the string to
end at the black hole. It would follow that a topologically 
stable string {\it cannot terminate on an extremal horizon}.
\footnote{Fundamental strings, which in many respects can be regarded as
infinitely thin global axion strings, do not possess, however, a core of 
unbroken symmetry which would be expelled. Therefore, they can end on 
extreme black holes--and more specifically, on D-branes 
\cite{polchinski}.}

Now, there have been a number of papers describing the pair creation of 
black holes with strings ending on them \cite{snap}. 
In order for the Euclidean instanton that 
mediates the process to be regular, the black holes must have (unconfined) 
charge, and be either extremal or close to extremality. 
This forces one to introduce, in addition to the massive gauge field
carried by the string, a (massless) $U(1)$ field to which the black hole
charge couples. Effectively, one works in a $U(1)^2$
theory of the same kind we have been discussing in this paper. 
But if, as we have argued, the string can not end at the extreme horizon, 
the corresponding instanton does not exist. This would imply
that a Nielsen-Olesen string could not snap by forming extreme black holes
at its ends. Thus, consideration of `realistic' strings seems to 
impose new selection rules on string snapping, of a 
sort somewhat different from those recently discussed in \cite{ag}.
A more detailed discussion of these results can be found in \cite{caes}.\\

The authors wish to express their thanks to Ana Ach{\'u}carro, Fay Dowker,
Ruth Gregory and Simon Ross.
A.C. was supported by NSF PHY94-07194 at ITP, Santa Barbara and 
by Pembroke College, Cambridge. J.M.A. A.-C. was supported by Wolfson College, 
University of Oxford. R.E. was partially supported by
a postdoctoral FPI fellowship (MEC-Spain) and by grant 
UPV 063.310-EB225/95. A.S. was supported by U.K.\ PPARC grant GR/L21488.


\end{document}